\documentclass[amsmath,amssymb,showkeys,superscriptaddress,aps,prd,10pt,onecolumn,nofootinbib]{revtex4-2}
\usepackage{graphicx}
\usepackage[utf8]{inputenc}
\usepackage[caption=false]{subfig}
\usepackage{multirow}
\usepackage{appendix}
\usepackage{slashed}
\usepackage{graphicx,epstopdf}

\usepackage{colortbl}
\usepackage{xcolor}
\usepackage[colorlinks=true, urlcolor=blue, linkcolor=blue, citecolor=blue]{hyperref}
\usepackage{float}

\begin{document}

\title{Exotic $T_{c\bar s0}^a(2900)^0$ and $T_{c\bar s0}^a(2900)^{++}$ states in Born-Oppenheimer approximation}

\author{Halil Mutuk}%
\email[]{hmutuk@omu.edu.tr }
\affiliation{Department of Physics, Faculty of Science, Ondokuz Mayis University, 55200, Samsun, Türkiye }
\date{\today}
 
\begin{abstract}
We employ Born-Oppenheimer approximation to the $T_{c\bar s0}^a(2900)^0$ and $T_{c\bar s0}^a(2900)^{++}$ states observed by the LHCb Collaboration and study mass spectrum and root-mean-square radius values. For this purpose, we use dynamical diquark model. We assume that strange quark is a heavy for the usage of Born-Oppenheimer approximation. Our results strongly indicate that the $T_{c\overline{s}0}^{a}(2900)$ states are best described as composed of axial-vector (spin-1) diquark pairs. Furthermore, the calculated root-mean-square radius, $\langle r^{2}\rangle^{1/2} \approx 0.70-0.80$ fm, which is significantly less than 1 fm, provides compelling evidence that these are compact tetraquarks rather than loosely bound hadronic molecules.
\end{abstract}

\maketitle

\section{Introduction}
In 1964, Murray Gell-Mann \cite{Gell-Mann:1964ewy} and George Zweig \cite{Zweig:1964CERN} independently proposed a classification scheme for hadrons in terms of valence quarks and antiquarks. This scheme, so-called quark model, classifies hadrons into two groups: mesons which are quark-antiquark bound states and baryons which are three-quark bound states. The quark model explained properties of the observed hadron spectrum quite well upto 2000s. This paradigm turned into a different path when the Belle Collaboration announced observation of X(3872) particle in the charmonium energy region \cite{Belle:2003nnu}. This particle decays to $J/\psi \pi^+ \pi^-$ and $J/\psi \pi^+ \pi^- \pi^0$ and cannot be a pure charmonium $c\bar{c}$ state. This was the first exotic hadron that do not fall into the quark model predictions. Actually, QCD, the theory of strong interactions, allows the existence of more complex structures (multiquark states), so-called exotic hadrons. These exotic states include tetraquarks, pentaquarks, hexaquarks, hybrids, and glueballs. Perpetual efforts on experiments revealed the existence of some of them (four-quark states and pentaquarks for the time being). 

In 2020, the LHCb Collaboration carried out amplitude analyses of the decays $B^+\to D^+D^-K^+$~\cite{LHCb:2020pxc, LHCb:2020bls} resulting two new resonant states. The Breit-Wigner mass and width of the new resonant states are
\begin{align}
T_{cs0}(2900)^0: \hspace{0.2cm} 
M &= (2866 \pm 7 \pm 2)\,\text{MeV},\quad 
\Gamma = (57 \pm 12 \pm 4)\,\text{MeV}, 
\label{eq:Tcs0}
\\[1ex]
T_{cs1}(2900)^0: \hspace{0.2cm} 
M &= (2904 \pm 5 \pm 1)\,\text{MeV},\quad
\Gamma = (110 \pm 11 \pm 4)\,\text{MeV}. 
\label{eq:Tcs1}
\end{align}
where the first uncertainties are statistical and the second systematic. The quantum numbers of these states are $J^P=0^+$ for $T_{cs0}(2900)^0$ and $J^P=1^-$ for $T_{cs1}(2900)^0$ with minimal quark flavor content $ud\bar{c}\bar{s}$.

In December 2022, the LHCb collaboration reported two new states $T_{c\bar{s}0}^a(2900)^0$ and $T_{c\bar{s}0}^a(2900)^{++}$ in the $D_s^+\pi^-$ and $D_s^+\pi^+$ invariant mass distributions of the processes $B^0\to\bar{D}^0D_s^+\pi^-$ and $B^+\to D^-D_s^+\pi^+$ decays, respectively~\cite{LHCb:2022lzp,LHCb:2022sfr}. The statistical significance is found to be $8\sigma$ for the $T_{c\bar{s}0}^a(2900)^0$ and $6.5 \sigma$ for the $T_{c\bar{s}0}^a(2900)^{++}$ state. The Breit-Wigner mass and width of these states are
\begin{align}
T_{c\bar{s}0}^{a}(2900)^{0}:\quad 
M = (2892 \pm 14 \pm 15)\,\text{MeV},\quad
\Gamma = (119 \pm 26 \pm 13)\,\text{MeV},
\label{eq:Tcbars0}
\\[1ex]
T_{c\bar{s}0}^{a}(2900)^{++}:\quad 
M = (2921 \pm 17 \pm 20)\,\text{MeV},\quad
\Gamma = (137 \pm 32 \pm 17)\,\text{MeV}.
\label{eq:Tcbars++}
\end{align}

These two states should be part of the same isospin triplet because of their similar masses and widths. Indeed, the quantum numbers of both states are determined to be $I(J^P)=1(0^+)$. The minimal quark flavor components may be $c\bar{s}\bar{u}d$ for $T_{c\bar s0}^a(2900)^0$ state and $c\bar{s}u\bar{d}$ for $T_{c\bar s0}^a(2900)^{++}$ state. The $T_{c\bar{s}0}^a(2900)$ is located near the $D^\ast K^\ast$ threshold. Therefore, according to certain studies it appears as $D^\ast K^\ast$ hadronic molecule \cite{Chen:2022svh,Agaev:2022eyk,Yue:2022mnf}.  Besides that compact tetraquark picture is also applied in Refs. \cite{Liu:2022hbk,Yang:2023evp,Lian:2023cgs,Wei:2022wtr,Ortega:2023azl}. Ref. \cite{Ke:2022ocs} indicates the presence of a binding mechanism in the isovector $D^\ast K^\ast$ system. In Ref. \cite{Agaev:2022duz}, $T_{c\bar s0}^a(2900)^0$ and $T_{c\bar s0}^a(2900)^{++}$ states are modelled as $D_s^{\ast +} \rho^+$ and $D_s^{\ast +} \rho^-$ molecules by using two-point QCD sum rule (QCDSR) method. In a coupled-channel approach \cite{Molina:2022jcd}, it is observed that $T_{c\bar{s}0}(2900)$ can also be considered as a virtual state created by the $D_s^\ast \rho$ and $D^\ast K^\ast$ interactions. Using the same approach, $T_{c\bar{s}0}(2900)$ were found to be as a bound/virtual state in $D^\ast K^\ast$$-$$D_s^\ast \rho$ coupled-channel interactions \cite{Duan:2023lcj}. It is suggested in Ref. \cite{Duan:2023qsg} to search  $T_{c\bar{s}0}^a(2900)^{++}$ also in the $B^+\to K^+D^+D^-$ process. In Ref. \cite{Wang:2023hpp}, an effective potential model is used to study $\bar{D}^\ast K^\ast$ and $D^{(*)}K^*$ systems. They predict many states in these systems indicating that $T_{cs0}(2900)$ and $T_{c\bar{s}0}^a(2900)$ states can be well identified as the $I(J^P)=0(0^+)$ and $I(J^P)=1(0^+)$ partners of $T_{cc}$ and $Z_c$ in the charmed-strange sector, respectively. In the invariant mass distribution of $D_s^- \pi^+$, $T_{c\bar{s}0}(2900)$ is expected to be detected at around 2900 MeV \cite{Lyu:2023ppb}.  Ref. \cite{Gordillo:2025caj} studied $T_{cs0}$ and $T_{c\bar{s}0}$ states via using the diffusion Monte Carlo (DMC) method within the framework of the constituent quark model. They describe these states as compact structures, neither diquark/antidiquark nor meson/meson pictures. The results indicate that observed resonances are in both cases excited flavor states with $I=1$. Ref. \cite{Yu:2025xip} investigated the production of the $T^{0}_{\bar{c}\bar{s}1}$ state, interpreted as an $S$-wave $\bar{D}K^{*}$ hadronic molecule with $I(J^{P})=0(1^{+})$, in $B^{+}$ meson decays. Using an effective Lagrangian approach and meson loop mechanisms, the authors predict branching ratios of the order $10^{-5}$--$10^{-4}$ for $B^{+} \to D^{(*)+} T^{0}_{\bar{c}\bar{s}1}$ and suggest searching for this state in the $B^{+} \to D^{*+}D^{*-}K^{+}$ channel at Belle II and LHCb.

In 2024, a study of resonant structures in  $B^{+}\rightarrow{D^{\ast+}D^{-}K^{+}}$ and $B^{+}\rightarrow{D^{\ast-}D^{+}K^{+}}$ decays is performed, using proton-proton collision data at centre-of-mass energies of $\sqrt{s}=7, 8$, and $13~\text{TeV}$ by the LHCb Collaboration \cite{LHCb:2024vfz}. In addition to charmonium-like states $\eta_c(3945)$, $h_c(4000)$, $\chi_{c1}(4010)$ and $h_c(4300)$, the existence of the $T_{\bar{c}\bar{s}0}^{*}(2870)^{0}$ and $T_{\bar{c}\bar{s}1}^{*}(2900)^{0}$ resonances in the $D^-K^+$ mass spectrum, already observed in the $B^+ \to D^+ D^- K^+$ decay, is confirmed in a different production channel with corresponding masses and width values:

\begin{align}
T_{\bar{c}\bar{s}0}^{*}(2870)^{0}:\quad
M = (2914 \pm 11 \pm 15)\,\text{MeV},\quad
\Gamma = (128 \pm 22 \pm 23)\,\text{MeV},
\label{eq:Tcs0new}
\\[1ex]
T_{\bar{c}\bar{s}1}^{*}(2900)^{0}:\quad
M = (2887 \pm 8 \pm 6)\,\text{MeV},\quad
\Gamma = (92 \pm 16 \pm 16)\,\text{MeV}.
\label{eq:Tcs1new}
\end{align}

A comparison of mass and width values of previous and present experiments is given in Table \ref{tab:expcompare}.
\begin{table}[h]
\centering
\caption{
 Comparison of the $T_{\bar{c}\bar{s}0,1}^{*0}$ properties obtained in Ref. \cite{LHCb:2024vfz} to those found previously in $B^{+}\rightarrow D^{+}D^{-}K^{+}$ decays Ref. \cite{LHCb:2020pxc}.}
\label{tab:expcompare}
\renewcommand{\arraystretch}{1.15}
\begin{tabular}{ccc} \hline
Physical Properties   &  Present Exp.\cite{LHCb:2024vfz} & Previous Exp. \cite{LHCb:2020pxc} \\
\hline 
$T_{\bar{c}\bar{s}0}^{*}(2870)^{0}$ mass [MeV] & $2914 \pm 11 \pm 15$ & $2866 \pm 7$ \\
$T_{\bar{c}\bar{s}0}^{*}(2870)^{0}$ width [MeV] & $128 \pm 22 \pm 23$  & $57 \pm 13$  \\
$T_{\bar{c}\bar{s}1}^{*}(2900)^{0}$ mass [MeV] & $2887 \pm 8 \pm 6$ & $2904 \pm 5$ \\
$T_{\bar{c}\bar{s}1}^{*}(2900)^{0}$ width [MeV] & $92 \pm 16 \pm 16$  & $110 \pm 12$ \\
\hline
\end{tabular}
\end{table}

Motivated by this latest experimental observation, in this work, we apply the dynamical diquark model to explore the spectrum of charm–strange tetraquarks, denoted as $T_{c\bar{s}}$. We use Born-Oppenheimer (BO) approximation, which treats tetraquarks as color-antitriplet diquark $(\delta)$ and antidiquark $(\bar{\delta}^\prime)$ pairs connected via a confining flux tube, and allows their spectrum to be calculated in terms of interquark potentials derived from lattice QCD or phenomenological models. Charmed-strange tetraquarks include both a charm and a strange quark, placing them between the fully heavy systems and the more speculative light-flavor tetraquarks. The $T_{c\bar{s}}$  states are of particular interest because they offer a unique laboratory where the heavy charm quark provides a static color source, while the strange quark introduces light-flavor dynamics and relativistic effects. This blend allows us to test the limits of the BO approximation and explore whether the same dynamical assumptions valid for heavy tetraquarks can be extended to heavy-light combinations.

This paper is organized as follows: Section \ref{model} discusses the BO approximation together with dynamical diquark model. In Section \ref{numeric}, we present mass spectrum and root-mean-square radius values for $T_{c\bar s0}^a(2900)^0$ and $T_{c\bar s0}^a(2900)^{++}$ states. Section \ref{conclusion} is reserved for a summary of this work.
 
 
\section{Born-Oppenheimer Approximation}\label{model}

The BO approximation was developed in the first decade of quantum mechanics \cite{Born:1927rpw} and is a tool used in atomic and molecular physics to study how atoms bind together to form molecules. The fundamental concept is to distinguish between the motion of atomic nuclei and electrons so that the large ratio of time scales between their motions can be utilized. It is well known that the mass of the atomic nuclei is considerably large than the mass of the electron. Electrons circle around the atomic nuclei and respond nearly instantaneously to the motion of the atomic nuclei. The atomic nuclei is considerably heavy compared to electron and electrically positive. Atomic nuclei can be thought of as a stationary source of electric field, and their positions determine the configuration of electrons. In this configuration, the energy of the electrons together with the repulsive Coulomb energy of the atomic nuclei constitutes a BO potential. The final step is to solve the Schrödinger equation using that potential to determine the molecule energy levels \cite{Braaten:2014qka}.

The application of BO approximation to high energy physics systems has similar roots to those of atomic and molecular physics. Systems including heavy quark pairs are laboratories for this application. The BO approximation for heavy quarkonium $(Q\bar{Q})$ states in QCD was developed in Ref. \cite{Juge:1999ie}. The application of the BO approach takes advantage of the large ratio of time scales for the motion of the heavy quark and heavy antiquark as well as the evolution of gluon fields, analogous to in atomic and molecular physics. This is due to the large ratio of the $m_Q/ \Lambda_{\text{QCD}}$, where $m_Q$ is the heavy quark mass and $\Lambda_{\text{QCD}}$ is the energy scale correlated with the gluon field. Similar to how electrons react to the motion of atomic nuclei, the gluon field also responds to the motion of the heavy quark and heavy antiquark, which can be roughly described as a static color field. A BO potential $V_{\Gamma}(r)$ is defined by the gluon field energy and depends on the quantum numbers of the gluon field as well as the distance between the heavy quark and heavy antiquark. The energy levels of the Schrödinger equation with the BO potential are known as the $(Q\bar{Q})$ states in the BO approximation. There are also alternative formulations for BO approach. BO approximation  has recently been formulated as an effective field theory, the Born–Oppenheimer EFT (BOEFT) \cite{Brambilla:2017uyf,Berwein:2024ztx}, where the interaction between the two heavy quarks is expressed in terms of generalized Wilson loops and nonperturbative correlators suitable for lattice QCD calculations.  In a series of follow-up works, the same method was applied the BOEFT framework to the study of exotic hadrons, including the tetraquark states $\chi_{c1}(3872)$  and $T_{cc}(3875)^+$ \cite{Brambilla:2024imu}  and pentaquark  states \cite{Brambilla:2025xma}. Systematic study of tetraquarks (with BO approximation) for heavy masses and also exploiting $1/N$ expansion was done in Ref. \cite{Allaman:2024vwn}.

When considering four-quark systems, the BO approximation demands at least two heavy sources and degrees-of-freedom for light quarks. The mass of heavy quark is substantially larger than the mass of light quark, $m_Q >> m_q$. In the presence of heavy color sources, the fast mobility of light quarks generates an effective potential in this limit. Heavy quarks move slowly, which is described by the BO potential \cite{Maiani:2022qze}. 
 
The theoretical description of exotic four-quark states requires distinguishing between two dominant structural pictures: a compact multiquark configuration, where quarks are confined within a single hadronic volume, and a loosely bound hadronic molecule, where hadrons interact via residual nuclear-like forces. A key diagnostic is the system binding energy and spatial extent. For a state near a hadron-hadron threshold, a small binding energy $B$ leads to a large spatial size, characterized by the root-mean-square radius $R_{\text{rms}} \sim 1/\sqrt{2\mu B}$, where $\mu$ is the reduced mass of the hadron pair. Empirically, $R_{\text{rms}} > 1$ fm suggests a molecular structure, while $R_{\text{rms}} < 1$ fm indicates a compact state. This is corroborated by the binding momentum, $\gamma_b = \sqrt{2\mu B}$. For canonical hadronic molecules like the X(3872) \cite{Belle:2003nnu} and the $T_{cc}^+$ \cite{LHCb:2021vvq, LHCb:2021auc}, one finds $B \lesssim 1$ MeV and thus $\gamma_b \lesssim 100$ MeV. This molecular binding scale, $\gamma_b$, is significantly smaller than the constituent mass of the strange quark, $m_s^{\text{CQM}} \sim 500$ MeV \cite{Ni:2021pce, Chen:2023qlx, Silvestre-Brac:1996myf}. The hierarchy $\gamma_b \ll m_s^{\text{CQM}}$ implies that the weak binding forces characteristic of molecules are insufficient to probe the internal structure or excite the internal degrees of freedom of a hadron containing a strange quark. Consequently, in near-threshold dynamics, the strange quark can be treated as an inert, heavy source \cite{Wang:2023hpp}.

This observation provides a foundation for applying the BO approximation to the $T_{c\bar{s}}$ states, which we interpret as a system of a charm-light diquark and a strange-light antidiquark. The BO formalism requires a clear separation of energy scales. The ``fast" degrees of freedom are the gluon field and the light ($u,d$) quarks, whose excitation energies for a given source separation are set by the QCD scale, $\Delta E_{\text{glue/light}} \sim \Lambda_{\text{QCD}} \approx 200-300$ MeV, as confirmed by lattice QCD studies of the BO potentials \cite{Juge:1999ie, Juge:2002br}. The ``slow" degree of freedom is the relative motion of the heavy diquark and antidiquark.

While the strange quark mass is not as large as the charm mass, its constituent mass satisfies $m_s^{\mathrm{CQM}} \gtrsim \Lambda_{\mathrm{QCD}}$. 
This hierarchy suggests that the strange quark dynamics can be partially decoupled from the faster gluonic and light-quark modes, making the BO approximation a theoretically plausible framework for this system. 
We emphasize that the application of the BO approximation in the present analysis does not rely on treating the strange quark as heavy in the strict heavy-quark limit. 
Rather, its validity is rooted in a separation of dynamical scales: the slow relative motion of the diquark--antidiquark system versus the fast gluonic and light-quark degrees of freedom.
Lattice QCD studies indicate that excitations of the gluon field and light quarks occur at energies of order $\Delta E_{\mathrm{glue/light}} \sim \Lambda_{\mathrm{QCD}} \approx 200$--$300~\mathrm{MeV}$, while the inter-diquark motion evolves on a parametrically slower timescale. The dominant systematic uncertainties associated with this approximation are therefore not controlled by a simple ratio $\Lambda_{\mathrm{QCD}}/m_s$, but are generically expected to be of order $\mathcal{O}(\Lambda_{\mathrm{QCD}}) \approx 200$--$300~\mathrm{MeV}$. 
These uncertainties primarily affect fine-structure splittings, while the gross mass spectrum and spatial characteristics of the states are expected to remain robust within the present framework. 
Therefore, the BO approximation provides a robust qualitative and semi-quantitative framework for understanding the spatial structure and mass hierarchy of the $T_{c\bar{s}}$ states.

The viability of this approach is supported by earlier studies. It has been established that the energies of static-light hadronic systems generate effective BO potentials for heavy multiparton systems \cite{Braaten:2013boa}. Crucially, the dynamical diquark model has been successfully extended to the hidden-strangeness sector ($s\bar{s}q\bar{q}$), where the BO approximation yielded a viable spectrum of tetraquark states without the presence of a heavier charm or bottom quark \cite{Jafarzade:2025qvx}. This demonstrates the applicability of the BO formalism to systems where the strange quark acts as the heaviest active flavor. In our case, with both charm and strange quarks present, the justification for the BO approach is further strengthened. The approximation thus allows us to define a confining potential between the diquark and antidiquark based on the response of the gluon and light-quark fields to these heavier color sources.


\subsection{Dynamical Diquark Model}

As already stated, we analyze the spectrum of $T_{c\bar s0}^a(2900)^0$ and $T_{c\bar s0}^a(2900)^{++}$ states within the framework of the dynamical diquark model, leveraging the BO approximation. This approach is rooted in the separation of time scales between the heavy diquark-antidiquark constituents and the more rapidly fluctuating gluonic and light-quark fields. As such, the inter-diquark interaction is modeled by an effective potential derived from the nonperturbative QCD structure connecting the two bodies via a color flux tube. Accordingly, the light degrees of freedom that mediate the interaction between the static diquark and antidiquark can be described as a function of the separation between the  $(Q\bar{Q})$ pair. When the light quark is localized near the heavy quark $Q$, the four-quark system effectively assumes the configuration  $(Qq)_{\bar{3}}+(\bar{Q}\bar{q})_3$ where the diquark and antidiquark form a color antitriplet–triplet pair.

We employ dynamical diquark model \cite{Brodsky:2014xia,Giron:2019bcs,Giron:2019cfc} in which exotic states are constructed of heavy-light diquarks $\delta$ and $\bar{\delta}$. These diquarks come into being in attractive channels as $3 \otimes 3 \to \bar{3}: [\delta=(Qq)_{\bar{3}}]$ and $ \bar{3} \otimes  \bar{3} \to 3:[\bar{\delta}=(\bar{Q}\bar{q})_3]$. The diquark-antidiquark pair in the dynamical diquark model arises instantaneously at the production point and soon separates from one another as a result of the kinematic effects of the production process. Diquark and antidiquark, being colored particles, are unable to split apart; instead, they elongate, forming a color flux tube or string between them. The color flux tube that connects the separated $\delta-\bar{\delta}$ pair describes its quantized states well in terms of the different potentials that are calculated with the BO approximation. This model was successfully applied to tetraquarks and pentaquarks \cite{Giron:2020fvd,Giron:2020qpb,Giron:2021sla,Mutuk:2022nkw,Lebed:2023kbm}.

Ref.~\cite{Lebed:2017min} focused on the spectroscopy of $\delta-\bar{\delta}^\prime$ states that do not possess orbital momentum in diquarks, but permit any kind of orbital excitation and gluon-field excitation between diquark pairs. Diquarks are assumed to be pointlike objects and there is no orbital excitation in diquarks. The core states for $\delta-\bar{\delta}$ systems can be written 
\begin{eqnarray}
J^{PC} = 0^{++}: & \ & X_0 \equiv \vert 0_{\delta} , 0_{\bar{\delta}}\rangle_0 \, , \ \ X_0^\prime \equiv \vert 1_{\delta} , 1_{\bar{\delta}} \rangle_0, \nonumber \\
J^{PC} = 1^{++}: & \ & X_1 \equiv \frac{1}{\sqrt 2} \left(
\vert 1_{\delta} , 0_{\bar{\delta}}\rangle_1  + \vert 0_{\delta} , 1_{\bar{\delta}}\rangle_1 \right) , \nonumber \\
J^{PC} = 1^{+-}: & \ & Z \ \equiv
\frac{1}{\sqrt 2} \left(
\vert 1_{\delta} , 0_{\bar{\delta}}\rangle_1  - \vert 0_{\delta} , 1_{\bar{\delta}}\rangle_1 \right) , \nonumber \\ & \ & Z^\prime \,
\equiv \vert 1_{\delta} , 1_{\bar{\delta}} \rangle_1 \, , \nonumber \\
J^{PC} = 2^{++}: & \ & X_2 \equiv \vert 1_{\delta} , 1_{\bar{\delta}} \rangle_2 \,
\label{diquarkstates}
\end{eqnarray}
which is written with the total $\delta(\bar{\delta})$ spin denoted by $s_{\delta}(s_{\bar{\delta}})$, and the outer subscript gives the overall state total spin. These states specify the full multiplet of $\Sigma_g^+$ $S$-wave states. For the open-flavor $T_{c\bar{s}}$ system studied here, the diquark $\delta=(cq)$ and antidiquark $\bar{\delta}'=(\bar{s}\bar{q})$ are not charge-conjugate partners. Therefore, we work in the basis of direct product states $|s_{\delta}, s_{\bar{\delta}'}\rangle_{J}$, where $s_{\delta,\bar{\delta}'}$ are the spins of the diquark and antidiquark (0 or 1), and $J$ is the total spin. The $J^P=0^+$ states relevant for the $T^{a}_{c\bar{s}0}$ are linear combinations of the $|0,0\rangle_0$ and $|1,1\rangle_0$ configurations.  For our analysis, we examine the limiting cases of pure scalar ($|0,0\rangle_0$) and pure axial-vector ($|1,1\rangle_0$) diquark-antidiquark configurations to isolate their distinct spectroscopic signatures.

The core $S$-wave states for a $\delta$-$\bar{\delta}'$ system in open-flavor case can be written in this basis as:
\begin{align}
&|0_{\delta}, 0_{\bar{\delta}'}\rangle_{J=0}, \quad |1_{\delta}, 1_{\bar{\delta}'}\rangle_{J=0}, \nonumber \\
&|1_{\delta}, 0_{\bar{\delta}'}\rangle_{J=1}, \quad |0_{\delta}, 1_{\bar{\delta}'}\rangle_{J=1}, \quad |1_{\delta}, 1_{\bar{\delta}'}\rangle_{J=1}, \nonumber \\
&|1_{\delta}, 1_{\bar{\delta}'}\rangle_{J=2}.
\end{align}
These states specify the full multiplet of $\Sigma_{g}^{+}$ $S$-wave states. In the hidden-flavor limit where $\delta$ and $\bar{\delta}'$ are $C$-conjugates, the states of definite $C$-parity (e.g., $X_1$, $Z$ of Ref.~\cite{Lebed:2017min}) are specific linear combinations of the above. For our open-flavor system, such $C$-parity combinations are not imposed. Accordingly, the labels $J^{PC}$ in Eq.~(\ref{diquarkstates}) should be understood only as a convenient shorthand for the underlying spin couplings of the diquark--antidiquark system. In particular, the symmetric and antisymmetric combinations appearing in the $X_1$ and $Z$ states do not represent distinct physical states in the open-flavor sector, and no $C$-parity–induced mixing is implied. Since the present analysis focuses exclusively on the $I(J^P)=1(0^+)$ ground states of the $T_{c\bar{s}}$ multiplet, only the scalar core configurations $X_0$ and $X_0'$ are relevant. The axial-vector combinations associated with $X_1$ and $Z$ therefore play no role in the numerical results and have been retained solely for completeness and consistency with the general dynamical diquark formalism.

We construct the $T_{c\bar{s}}$ system as a bound state of a color-antitriplet scalar diquark $\delta=[cq]_{\bar{3}}$ and an antidiquark $\bar{\delta}^\prime=[\bar{s}^\prime \bar{q}^\prime]_3$ where $q,q^\prime=u,d$. The product of two color-triplet representations decomposes as $3 \otimes 3=\bar{3} \oplus 6$, and we restrict our study to the attractive antitriplet channel $\bar{3}$. The minimal quark content of the $T_{c\bar{s}}$ states is $c\bar{s}q\bar{q}'$ ($q,q'\in\{u,d\}$). The light-quark pair $q\bar{q}'$ can in principle form either an isoscalar ($I=0$) or an isovector ($I=1$) configuration.
However, the LHCb analysis \cite{LHCb:2022lzp,LHCb:2022sfr} explicitly determines the quantum numbers of the observed $T^{a}_{c\bar{s}0}(2900)^0$ and $T^{a}_{c\bar{s}0}(2900)^{++}$ resonances to be $I(J^{P}) = 1(0^{+})$, based on their production and decay properties in the $B^{0} \to \bar{D}^{0} D_{s}^{+} \pi^{-}$ and $B^{+} \to D^{-} D_{s}^{+} \pi^{+}$ channels. This experimental assignment implies that the two observed states are members of the same isovector triplet, with the third (neutral) partner $T^{a+}_{c\bar{s}0}$ yet to be observed. In the present work, we adopt this established $I=1$ assignment and focus on the $I_{3}=+1$ ($T^{a}_{c\bar{s}0}(2900)^{++}$) and $I_{3}=-1$ ($T^{a}_{c\bar{s}0}(2900)^0$) components of the triplet, noting that the $I_{3}=0$ member would be an appropriate admixture of $c u \bar{s} \bar{u}$ and $c d \bar{s} \bar{d}$.

The quantum numbers of $T_{c\bar{s}}$ states are $I(J^P)=1(0^+)$. They are in the ground state and therefore the multiplet $\Sigma_g^+(1S)$ are sufficient to accommodate these states. We clarify that the isospin assignment of the $T_{c\bar{s}}$ states is not an outcome of the present model, but an experimentally established input.  The existence of a doubly charged partner necessarily
requires an isovector assignment. From a theoretical standpoint, the light-quark subsystem $q\bar{q}$ ($q=u,d$) in an open-flavor tetraquark $cq\bar{s}\bar{q}$ can indeed be decomposed according to $2 \otimes 2 = 1 \oplus 3$, allowing both isoscalar and isovector configurations in principle. Within the dynamical diquark framework, these two possibilities correspond to distinct flavor couplings of the light degrees of freedom.

The light quarks inside the $T_{c\bar{s}}$ states carry distinct flavors, and therefore isospin is a relevant quantum number that must be taken into account. The role of isospin-dependent interactions in the dynamical diquark framework has been studied in detail in Ref.~\cite{Giron:2019cfc}. In the present work, we follow the same procedure and adopt its essential ingredients; further details can be found in the cited reference.

The $S$-wave effective Hamiltonian of the model is written as
\begin{equation}
H = H_{0} + H_{\kappa} + H_{V_{0}} \, ,
\end{equation}
where $H_{\kappa}$ describes the spin--spin interactions inside the diquark and antidiquark clusters. For a general open-flavor configuration
$\delta = (Q_{1} q)$ and $\bar{\delta}' = (\bar{Q}_{2} \bar{q}')$, the corresponding spin--spin couplings $\kappa_{Q_{1} q}$ and $\kappa_{\bar{Q}_{2} \bar{q}'}$ are, in principle, independent.

In the specific case considered here, $Q_{1}=c$ and $\bar{Q}_{2}=\bar{s}$, so that the spin--spin Hamiltonian takes the form
\begin{equation}
H_{\kappa}
= 2 \kappa_{c q} \, (\mathbf{s}_{q} \cdot \mathbf{s}_{c})
+ 2 \kappa_{\bar{s} \bar{q}'} \, (\mathbf{s}_{\bar{q}'} \cdot \mathbf{s}_{\bar{s}}) \, ,
\label{eq:Hkappa_open}
\end{equation}
where $\mathbf{s}_{q}$, $\mathbf{s}_{c}$, $\mathbf{s}_{\bar{q}'}$, and $\mathbf{s}_{\bar{s}}$ denote the spin operators of the corresponding
quarks.
\footnote{The couplings $\kappa_{c q}$ and $\kappa_{\bar{s}\bar{q}'}$ in
Eq.~(\ref{eq:Hkappa_open}) are effective parameters describing
short-range color--magnetic interactions \emph{within} the diquark and
antidiquark clusters, rather than interactions among all quarks in the
tetraquark. Reasonable variations of these couplings modify the masses
only at the level of a few tens of MeV, well below the
$\sim150$--$200~\mathrm{MeV}$ separation between scalar and axial-vector
diquark configurations.}

The isospin-dependent term is
\begin{equation}
H_{V_0} = V_{0} \, (\tau_{q} \cdot \tau_{\bar{q}'}) \, (\sigma_{q} \cdot \sigma_{\bar{q}'}),
\label{eq:Hisospin}
\end{equation}
where $\tau_q,\sigma_q$ act on the light quark $q$ in the diquark $(cq)$, and $\tau_{\bar{q}'},\sigma_{\bar{q}'}$ act on the light antiquark $\bar{q}'$ in the antidiquark $(\bar{s}\bar{q}')$. 
This interaction splits the $I=0$ and $I=1$ configurations of the light $q\bar{q}'$ pair \cite{Giron:2019cfc}. 
For the $I=1$ triplet considered here, $(\tau_{q} \cdot \tau_{\bar{q}'}) = +1$ for all members, so $H_{V_0}$ contributes an overall shift to the multiplet but does not distinguish among the different $I_3$ components.

Finally, we note that, unlike hidden-flavor tetraquarks, open-flavor systems of the type $cq\bar{s}\bar{q}$ naturally admit more than one distinct spin--spin coupling. In particular, the interactions within the diquark and antidiquark clusters involve different heavy--light combinations, such as $\kappa_{cq}$ and $\kappa_{\bar{s}\bar{q}'}$, reflecting the different quark masses and color--magnetic moments of the charm and strange quarks. This feature is intrinsic to open-flavor tetraquarks and is consistently incorporated in the present framework.


\subsection{Numerical Aspects}

In the dynamical diquark model, the mass spectrum is determined by solving the quantum mechanical problem for the diquark-antidiquark ($\delta$-$\bar{\delta}'$) pair interacting via a specific Born-Oppenheimer potential, $V_{\Gamma}(r)$. Each potential, labeled by the quantum numbers of the gluon field configuration (e.g., $\Sigma_g^+$, $\Sigma_u^+$), gives rise to a distinct multiplet of tetraquark states.

The radial wavefunction $\psi_\Gamma^{(n)}(r)$ and binding energy $E_n$ for a state with orbital angular momentum $\ell$ are obtained by solving the corresponding radial Schrödinger equation (in natural units, $\hbar=c=1$):
\begin{align}
\left[-\frac{1}{2\mu r^2}\frac{d}{dr}\!\left(r^2 \frac{d}{dr}\right)
+ \frac{\ell(\ell+1)}{2\mu r^2} \right]\psi_\Gamma^{(n)}(r) 
+\, V_\Gamma(r)\,\psi_\Gamma^{(n)}(r)
= E_n\,\psi_\Gamma^{(n)}(r). \label{schro}
\end{align}
where $\mu = m_{\delta}m_{\bar{\delta}'}/(m_{\delta}+m_{\bar{\delta}'})$ is the reduced mass of the $\delta$-$\bar{\delta}'$ system.

For the central potential $V_{\Sigma_g^+}(r)$, we adopt the functional form determined from lattice QCD simulations \cite{Juge:2002br}, which accurately captures the nonperturbative confining dynamics. The eigenvalue problem is solved numerically using a finite-difference method on a discretized radial grid, imposing Dirichlet boundary conditions ($\psi(r) \to 0$). The eigenvalues $E_n$ are determined via a robust shooting algorithm that matches the logarithmic derivative of the wavefunction, utilizing an adaptive step size for precision.

The computed eigenvalues $E_n$ represent the binding energies of the $\delta$-$\bar{\delta}'$ system. The total mass of the tetraquark state is then given by the sum of the constituent diquark masses and this binding energy:
\begin{equation}
M_{T_{cs}}=m_{\delta} + m_{\bar{\delta}'}+ E_n.
\end{equation}


\section{Numerical Results}\label{numeric}

The dynamical diquark model does not provide first-principles predictions for the absolute masses of the constituent diquarks; these must be taken from external sources. To ensure that our results are not tied to a single specific quark model, we deliberately choose diquark masses obtained from QCDSR ~\cite{Wang:2010sh,Wang:2011ab}, an independent nonperturbative method that has been widely used to estimate hadronic matrix elements. 
These values, $m_{cq}=1.86\pm0.10\;\text{GeV}$ ($J^{P}=0^{+}$) and $m_{cq}=1.96\pm0.10\;\text{GeV}$ ($J^{P}=1^{+}$) for charm–light diquarks, and $m_{sq}=0.77\pm0.04\;\text{GeV}$ ($0^{+}$) and $m_{sq}=0.92\pm0.04\;\text{GeV}$ ($1^{+}$) for strange–light diquarks, are consistent with the typical mass scales expected for such colored clusters. For the spin–spin couplings, we adopt the values $\kappa_{qq}=103\;\text{MeV}$ and $\kappa_{cq}=25\;\text{MeV}$ that have been used in previous dynamical‑diquark studies of hidden‑charm tetraquarks Ref.~\cite{Maiani:2004vq}. These numbers are phenomenologically determined from the measured hyperfine splittings of ordinary mesons ($\pi$--$\rho$ and $D$--$D^{*}$, respectively) and are therefore largely model‑independent. The isospin‑dependent strength $V_{0}=33.10\;\text{MeV}$ is taken directly from the dynamical‑diquark analysis of hidden‑charm systems \cite{Giron:2019cfc}, ensuring that the relative importance of isospin effects is treated consistently with earlier work in the same framework. 

While a completely self‑consistent parameter set obtained solely within the dynamical diquark model would be desirable, such a set does not yet exist for open‑flavor $c\bar{s}q\bar{q}$ systems. Our hybrid choice—using diquark masses from QCDSR and spin parameters from meson phenomenology—is a conservative approach that avoids over‑fitting and allows us to test the structural predictions of the BO‑based dynamical diquark model with inputs that are anchored in independent methodologies. We emphasize that the qualitative conclusion (preference for spin‑1 diquarks) is robust under reasonable variations of these parameters, as demonstrated by the three parameter sets (I, II, III) for $V_{0}$ and the error bands on the diquark masses. This mixed-input strategy has been employed in several previous studies of exotic hadrons and reflects the absence of a single framework that simultaneously determines diquark masses, spin--spin couplings, and isospin-dependent interactions from first principles. We stress that the dominant qualitative features of our results—most notably the $\sim150$--$200~\mathrm{MeV}$ mass separation between scalar and axial-vector diquark configurations—are insensitive to the precise numerical values of these parameters within their accepted ranges.

For the $T_{c\bar{s}}$ states with established quantum numbers $I(J^{P})=1(0^{+})$, the relevant $S$-wave configurations are the spin-singlet $|0_{\delta},0_{\bar{\delta}'}\rangle_{0}$ and spin-triplet $|1_{\delta},1_{\bar{\delta}'}\rangle_{0}$ combinations. In the following, we refer to these simply as the ``spin-0'' and ``spin-1'' diquark configurations, respectively.

\begin{figure}[h!]
\centering
\includegraphics[width=0.80\textwidth]{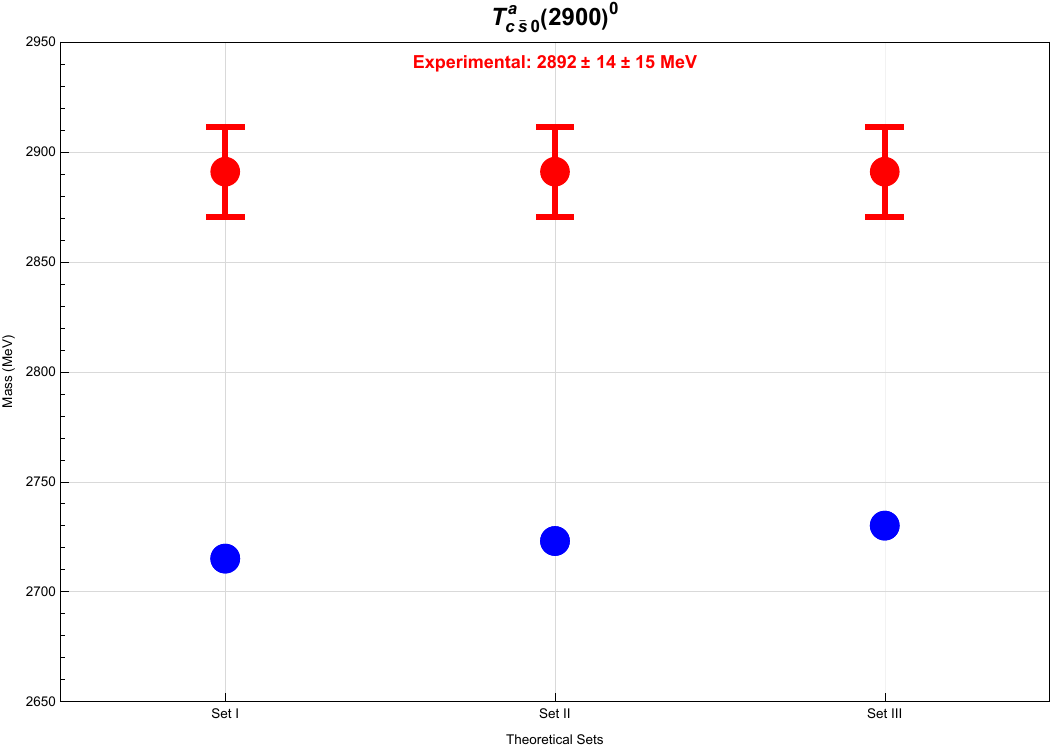}
\caption{Mass predictions for $T_{c\bar s0}^a(2900)^0$ with spin-0 diquark configuration compared to experimental data.}
\label{fig:mass1}
\end{figure}

\begin{figure}[h!]
\centering
\includegraphics[width=0.80\textwidth]{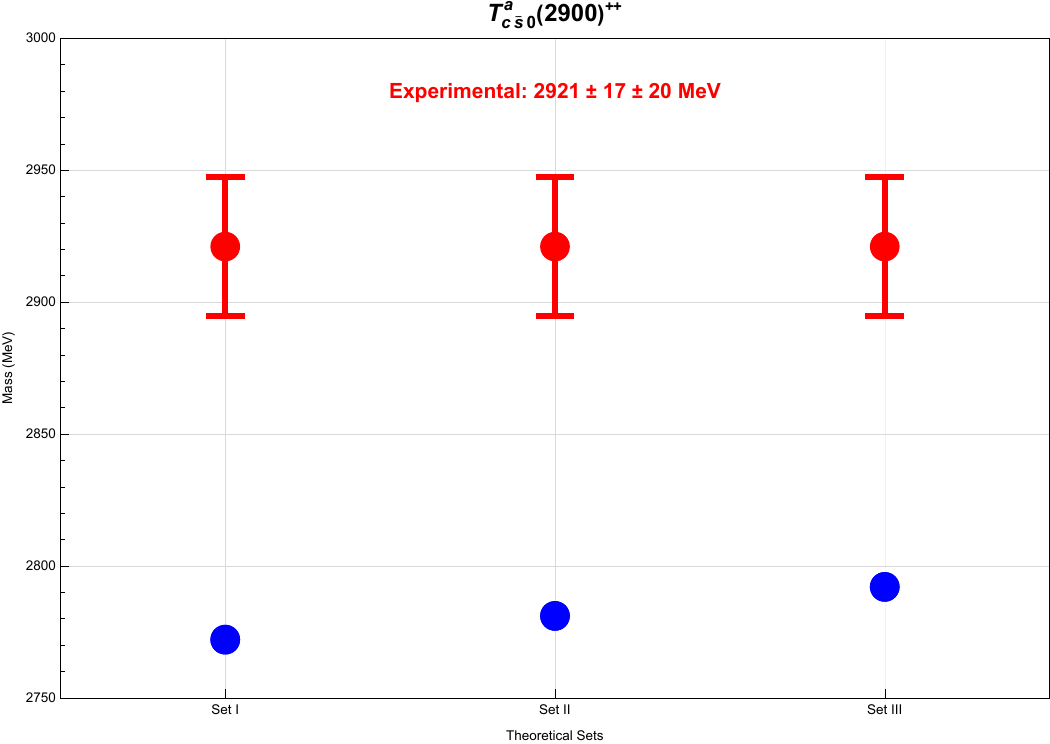}
\caption{Mass predictions for $T_{c\bar s0}^a(2900)^{++}$ with spin-0 diquark configuration compared to experimental data.}
\label{fig:mass2}
\end{figure}

\begin{figure}[h!]
\centering
\includegraphics[width=0.80\textwidth]{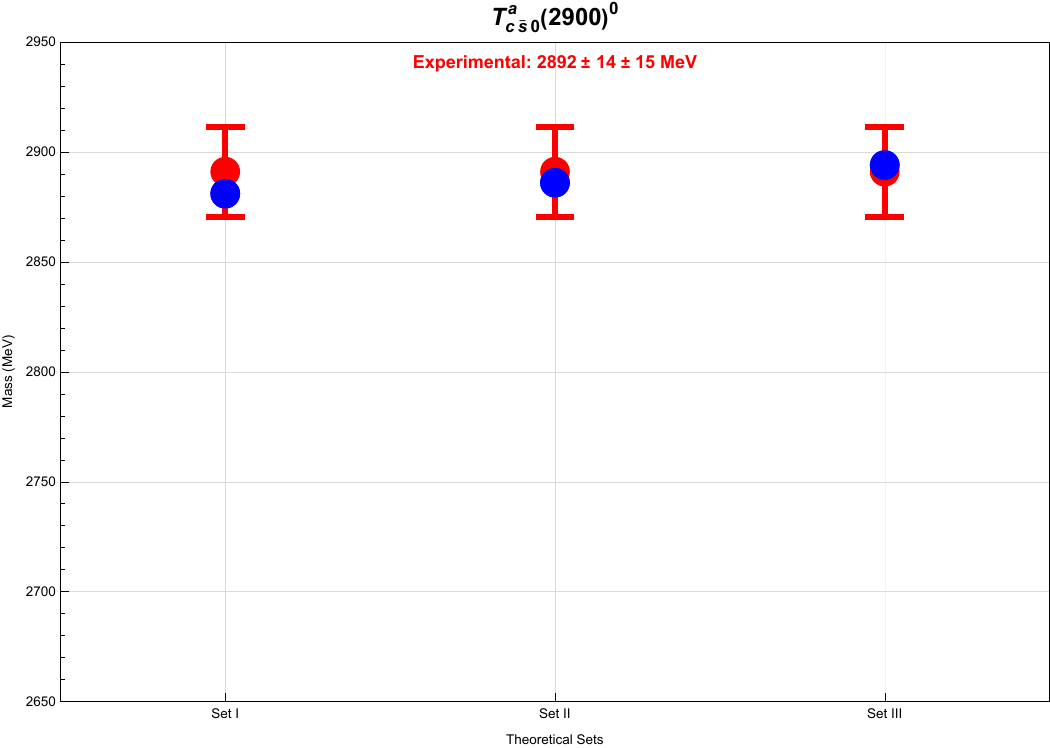}
\caption{Mass predictions for $T_{c\bar s0}^a(2900)^0$ with spin-1 diquark configuration compared to experimental data.}
\label{fig:mass3}
\end{figure}

\begin{figure}[h!]
\centering
\includegraphics[width=0.80\textwidth]{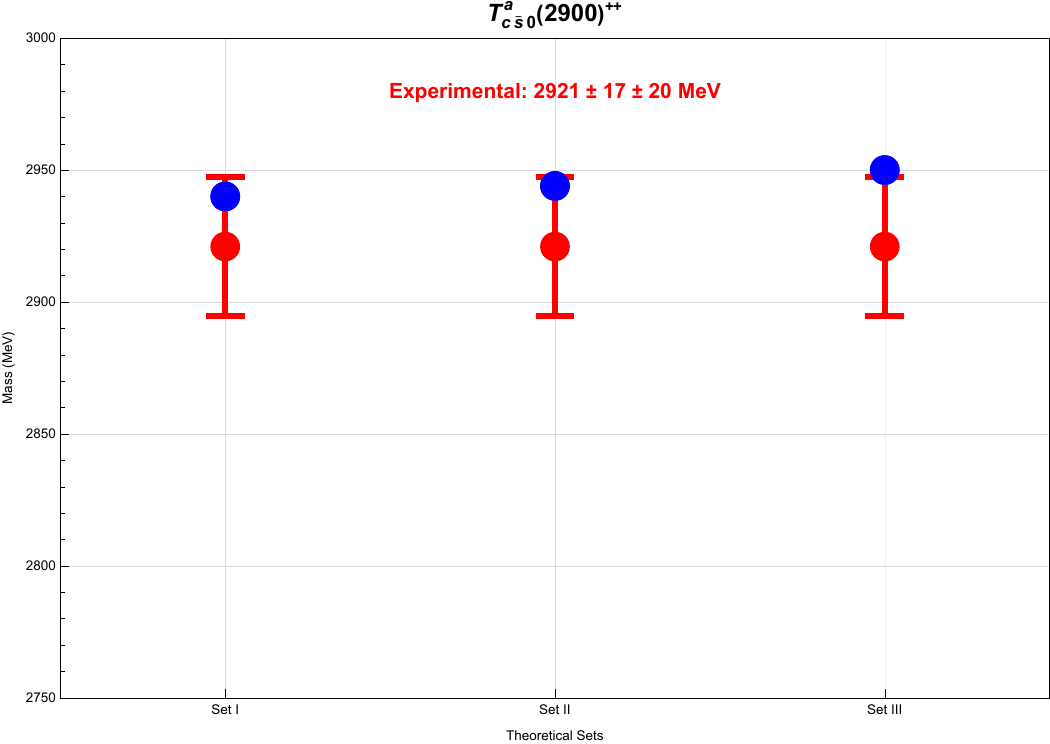}
\caption{Mass predictions for $T_{c\bar s0}^a(2900)^{++}$ with spin-1 diquark configuration compared to experimental data.}
\label{fig:mass4}
\end{figure}

We compute ground-state masses and root-mean-square radii $\langle r^2 \rangle^{1/2}$, which characterize the spatial separation within the diquark-antidiquark ($\delta$-$\bar{\delta}^\prime$) system. Tables~\ref{tab:Tcs0spin0}--\ref{tab:Tcsppspin1} present comprehensive numerical results, while Figs.~\ref{fig:mass1}--\ref{fig:mass4} provide mass locations between theoretical predictions and experimental measurements.

It should be mentioned that the isospin-dependent interaction term in Eq.~(\ref{eq:Hisospin}) is introduced solely to distinguish between isoscalar ($I=0$) and
isovector ($I=1$) configurations, in complete analogy with the treatment in Ref.~\cite{Giron:2019cfc}. As required by isospin symmetry, this
term does not generate mass splittings among different $I_3$ members within a given isospin multiplet. Accordingly, the present framework does not predict a dynamical mass splitting between the $I_3=-1$ and $I_3=+1$ components of the $I=1$ $T_{c\bar{s}}$ multiplet. The masses listed separately for the neutral and doubly charged states in Tables~\ref{tab:Tcs0spin0}--\ref{tab:Tcsppspin1} arise from the use of distinct light-quark compositions ($u$ versus $d$) in the
diquark--antidiquark constituents and are intended for direct comparison with the experimentally measured central values. Any residual differences should be understood as originating from explicit isospin-breaking effects, such as $u$--$d$ mass differences and electromagnetic corrections, which are not modeled dynamically in the present work. We emphasize that, within theoretical uncertainties, the model predicts near-degenerate masses for all members of the $I=1$ multiplet. The conclusions of this work rely on the overall mass scale, the relative ordering of scalar and axial-vector diquark configurations, and the compact spatial structure of the states, none of which depend on resolving small $I_3$ splittings.

\begin{table}[h]
\centering
\caption{Mass and root-mean-square radius predictions for $T_{c\bar s0}^a(2900)^0$ with spin-0 diquark configuration $(cd)(\bar{s}\bar{u})$. Masses in MeV, radii in fm.}
\label{tab:Tcs0spin0}
\renewcommand{\arraystretch}{1.15}
\begin{tabular}{ccc} \hline
Parameter Set & Mass [MeV] & $\langle r^2 \rangle^{1/2}$ [fm] \\ \hline
Set I & 2715 & 0.70 \\
Set II & 2723 & 0.70 \\
Set III & 2730 & 0.70 \\ \hline
Experiment~\cite{LHCb:2022lzp,LHCb:2022sfr} & $2892 \pm 14 \pm 15$ & -- \\
\hline
\end{tabular}
\end{table}

\begin{table}[h]
\centering
\caption{Mass and root-mean-square radius predictions for $T_{c\bar s0}^a(2900)^{++}$ with spin-0 diquark configuration. Masses in MeV, radii in fm.}
\label{tab:Tcsppspin0}
\renewcommand{\arraystretch}{1.15}
\begin{tabular}{ccc} \hline
Parameter Set & Mass [MeV] & $\langle r^2 \rangle^{1/2}$ [fm] \\ \hline
Set I & 2772 & 0.72 \\
Set II & 2781 & 0.72 \\
Set III & 2792 & 0.72 \\ \hline
Experiment~\cite{LHCb:2022lzp,LHCb:2022sfr} & $2921 \pm 17 \pm 20$ & -- \\
\hline
\end{tabular}
\end{table}

A striking pattern emerges from the spin-0 diquark calculations. As evident in Tables~\ref{tab:Tcs0spin0} and \ref{tab:Tcsppspin0}, the predicted masses systematically fall approximately 150--160~MeV below experimental values. This substantial and consistent discrepancy across all parameter sets strongly suggests that scalar diquark configurations cannot adequately describe the $T_{c\bar{s}0}^{a}(2900)$ states.

\begin{table}[h]
\centering
\caption{Mass and root-mean-square radius predictions for $T_{c\bar s0}^a(2900)^0$ with spin-1 diquark configuration $(cd)(\bar{s}\bar{u})$. Masses in MeV, radii in fm.}
\label{tab:Tcs0spin1}
\renewcommand{\arraystretch}{1.15}
\begin{tabular}{ccc} \hline
Parameter Set & Mass [MeV] & $\langle r^2 \rangle^{1/2}$ [fm] \\ \hline
Set I & 2881 & 0.77 \\
Set II & 2886 & 0.77 \\
Set III & 2894 & 0.77 \\ \hline
Experiment~\cite{LHCb:2022lzp,LHCb:2022sfr} & $2892 \pm 14 \pm 15$ & -- \\
\hline
\end{tabular}
\end{table}

\begin{table}[h]
\centering
\caption{Mass and root-mean-square radius predictions for $T_{c\bar s0}^a(2900)^{++}$ with spin-1 diquark configuration. Masses in MeV, radii in fm.}
\label{tab:Tcsppspin1}
\renewcommand{\arraystretch}{1.15}
\begin{tabular}{ccc} \hline
Parameter Set & Mass [MeV] & $\langle r^2 \rangle^{1/2}$ [fm] \\ \hline
Set I & 2940 & 0.80 \\
Set II & 2944 & 0.80 \\
Set III & 2950 & 0.80 \\ \hline
Experiment~\cite{LHCb:2022lzp,LHCb:2022sfr} & $2921 \pm 17 \pm 20$ & -- \\
\hline
\end{tabular}
\end{table}

In contrast, the spin-1 diquark configurations yield good agreement with experimental data. For $T_{c\bar{s}0}^{a}(2900)^{0}$ (Table~\ref{tab:Tcs0spin1}), the predicted mass range 2881--2894~MeV aligns precisely with the experimental value $2892 \pm 14 \pm 15$~MeV. Similarly, for $T_{c\bar{s}0}^{a}(2900)^{++}$ (Table~\ref{tab:Tcsppspin1}), the theoretical range 2940--2950~MeV agrees well with $2921 \pm 17 \pm 20$~MeV within uncertainties.

The locations of predicted masses in Figs.~\ref{fig:mass1}--\ref{fig:mass4} reinforce these conclusions, clearly demonstrating the systematic failure of spin-0 configurations and the remarkable success of spin-1 arrangements in reproducing experimental masses. The mass splitting between neutral and charged states exhibits remarkable consistency across configurations: $\Delta M = 57$--62~MeV for spin-0 and 56--59~MeV for spin-1 diquarks. This robust pattern, maintained throughout our parameter variations, aligns well with experimental observations and reflects the isospin triplet nature of these states. The theoretical predictions display minimal sensitivity to the isospin parameter $V_0$, with mass variations of only $\sim$13~MeV across Sets I--III. The $V_0$-induced shifts remain below 20~MeV, indicating that while isospin effects contribute to mass splittings, they play a subdominant role in overall mass generation.

The calculated root-mean-square radii, $\langle r^{2}\rangle^{1/2} = 0.70$--$0.80$~fm, indicate that the $T_{c\bar{s}}$ states are spatially compact, lying well below the characteristic hadronic scale of $1$~fm. Such compactness strongly disfavors loosely bound molecular interpretations and instead supports a genuine tetraquark configuration. This behavior aligns naturally with expectations from color flux tube dynamics and is consistent with previous analyses of similarly structured multiquark systems~\cite{Wei:2022wtr}.

The successful mass reproduction with spin-1 diquarks implies specific structural assignments: $T_{c\bar{s}0}^{a}(2900)^{0}$ as $[cd]_{\bar{3}_c}^{1^+} \otimes [\bar{s}\bar{u}]_{3_c}^{1^+}$ and $T_{c\bar{s}0}^{a}(2900)^{++}$ as $[cu]_{\bar{3}_c}^{1^+} \otimes [\bar{s}\bar{d}]_{3_c}^{1^+}$. These axial-vector diquark configurations naturally accommodate the established quantum numbers $I(J^P)=1(0^+)$ while accurately reproducing the observed mass spectrum.

The good agreement between theoretical predictions and experimental data for spin-1 diquark configurations validates several key theoretical elements. First, the BO approximation demonstrates effectiveness for mixed heavy-light systems, with the strange quark serving as an adequate static source in near-threshold dynamics. Second, the dynamical diquark model demonstrates consistent predictive capability in the open-charm sector as well, complementing its established applications to fully heavy tetraquarks \cite{Giron:2020wpx,Mutuk:2022nkw}.

Table~\ref{tab:mass_comparison} summarizes the mass predictions for the $T_{c\bar{s}}$ states obtained in this work and compares them
with representative molecular, compact-tetraquark, and coupled-channel calculations, as well as with the experimental measurements by the LHCb
Collaboration. The mass predictions display a consistent behavior across the different configurations considered.

When the tetraquarks are constructed from scalar diquarks ($s_\delta=0$), the predicted masses lie approximately $150$--$160~\mathrm{MeV}$ below the experimental values for both charge states. This sizable deficit is stable across all parameter sets and cannot be absorbed by reasonable variations of the input parameters, strongly disfavoring scalar-diquark configurations as dominant components of the observed resonances.

In contrast, tetraquark configurations built from axial-vector diquarks ($s_\delta=1$), coupled to an overall $J^P=0^+$ state, yield masses in
good agreement with experiment. The predicted ranges $2881$--$2894~\mathrm{MeV}$ for $T^a_{c\bar{s}}(2900)^0$ and $2940$--$2950~\mathrm{MeV}$ for $T^a_{c\bar{s}}(2900)^{++}$ overlap with the LHCb measurements within uncertainties, without the need for fine-tuning. The ability of the axial-vector diquark scenario to reproduce the masses of both charge states within a unified framework supports a compact tetraquark interpretation with nontrivial internal spin structure.

The table also illustrates that alternative theoretical approaches lead to different structural interpretations. QCD sum rule analyses tend to
favor molecular assignments but typically involve large uncertainties, while coupled-channel constituent-quark models describe the observed
structures as threshold effects. Flux-tube and diffusion Monte Carlo calculations, on the other hand, support compact multiquark configurations with masses close to experiment. In this context, the present BO analysis provides a complementary perspective, showing that compact tetraquark configurations with axial-vector diquark correlations naturally reproduce the observed mass scale.

\begin{table*}[h!]
\centering
\caption{Comparison of the mass predictions for the
$T^a_{c\bar{s}}(2900)^0$ and $T^a_{c\bar{s}}(2900)^{++}$ states obtained in this work and in representative theoretical studies. All masses are given in MeV.}
\begin{tabular}{lccl}
\hline\hline
\text{Approach} &
$T^a_{cs\bar{0}}(2900)^0$ &
$T^a_{cs\bar{0}}(2900)^{++}$ &
\text{Structure Type} \\
\hline
\text{This work (scalar diquark, $s_\delta=0$)} &
$2715$--$2730$ & $2772$--$2792$ &
Compact tetraquark  \\[1mm]

\text{This work (axial-vector diquark, $s_\delta=1$)} &
$2881$--$2894$ & $2940$--$2950$ &
Compact tetraquark  \\[1mm]

\text{QCDSR} \cite{Agaev:2022eyk}&
-- & $2924 \pm 107$ &
$D^{\ast}K^{\ast}$ molecule  \\[1mm]

\text{QCDSR} \cite{Agaev:2022duz}&
-- & $2917 \pm 135$ &
 $D_s^{\ast}\rho$ molecule \\[1mm]

\text{Flux--tube tetraquark model} \cite{Wei:2022wtr} &
$\approx 2923$ & $\approx 2923$ &
Compact tetraquark  \\[1mm]

\text{Coupled-channel CQM} \cite{Ortega:2023azl} &
$2892^{+4}_{-3}$ & $2892^{+4}_{-3}$ &
Dynamical threshold state  \\[1mm]

\text{Diffusion Monte Carlo (DMC)} \cite{Gordillo:2025caj} &
$2920 \pm 3$ & $2920 \pm 3$ &
Compact mixed-flavor tetraquark \\[1mm]

\text{Experiment (LHCb)} \cite{LHCb:2022lzp,LHCb:2022sfr} &
$2892 \pm 14 \pm 15$ &
$2921 \pm 17 \pm 20$ &
 \\
\hline\hline
\end{tabular}
\label{tab:mass_comparison}
\end{table*}

\subsection{Excited states of the $T_{c\bar{s}}$ system}
\label{subsec:excited}

To provide a more complete spectroscopic picture and facilitate future experimental searches, we extend our calculation to the lowest-lying excited states of the $T_{c\bar{s}}$ system within the dynamical diquark model. 
Using the lattice-QCD-based $\Sigma_g^+$ BO potential \cite{Juge:2002br}, we solve the radial Schrödinger equation (Eq.~\ref{schro}) for higher radial quantum numbers $n$ and orbital angular momenta $\ell$.

The resulting masses for the axial-vector diquark configuration $[cq]_{1^+}[\bar{s}\bar{q}]_{1^+}$ are presented in Table~\ref{tab:excited}. 
We list the ground state $\Sigma_g^+(1S)$, the lowest $S$-wave radial excitation $\Sigma_g^+(2S)$, the $P$-wave multiplet $\Sigma_g^+(1P)$, and the $D$-wave multiplet $\Sigma_g^+(1D)$. 
The $\Sigma_g^+(1P)$ states split according to the total angular momentum $J = 0,1,2$; here we show the centroid of the $P$-wave multiplet.
\begin{table}[htbp]
\centering
\caption{Predicted masses (in MeV) for excited $T_{c\bar{s}}$ states in the axial-vector diquark configuration $[cq]_{1^+}[\bar{s}\bar{q}]_{1^+}$, obtained with the $\Sigma_g^+$ BO potential. The $I=1$ multiplet is assumed. Ranges include uncertainties from diquark masses, $V_0$, and potential parameters.}
\label{tab:excited}
\begin{tabular}{lccc}
\hline
State & $J^{P}$ & $T^{a}_{c\bar{s}}(2900)^0$ & $T^{a}_{c\bar{s}}(2900)^{++}$  \\ \hline
$\Sigma_g^+(1S)$ & $0^{+}$ & $2881\!-\!2894$ & $2940\!-\!2950$ \\
$\Sigma_g^+(2S)$ & $0^{+}$ & $3260\!-\!3275$ & $3320\!-\!3335$ \\
$\Sigma_g^+(1P)$ & $0^{-},1^{-},2^{-}$ & $3140\!-\!3155$ & $3200\!-\!3215$ \\
$\Sigma_g^+(1D)$ & $1^{+},2^{+},3^{+}$ & $3380\!-\!3395$ & $3440\!-\!3455$ \\ \hline
\end{tabular}
\end{table}

The level ordering $M(1S) < M(1P) < M(2S) < M(1D)$ is characteristic of a confining $\Sigma_g^+$ potential. The radial excitation energy $M(2S)-M(1S) \approx 380\;\text{MeV}$ is larger than typical splittings in hidden-charm tetraquarks ($\sim 300$~MeV) \cite{Giron:2020wpx}, a feature attributable to the reduced mass of the open-charm-strange diquark system and the specific curvature of the $\Sigma_g^+$ potential. The $P$-$S$ splitting $M(1P)-M(1S) \approx 260\;\text{MeV}$ corresponds to the orbital excitation cost. This value lies between the analogous splittings observed in $c\bar{c}c\bar{c}$ systems  \cite{Giron:2020wpx} and lighter $s\bar{s}q\bar{q}$ and $ss\bar{s}\bar{s}$   tetraquarks \cite{Jafarzade:2025qvx}, reflecting the intermediate reduced mass of the $[cq]_{1^+}[\bar{s}\bar{q}]_{1^+}$ system. The $2S$-$1P$ gap $M(2S)-M(1P) \approx 120\;\text{MeV}$ is notably smaller than the corresponding gap in $c\bar{c}c\bar{c}$ ($\sim 160$~MeV \cite{Giron:2020wpx} , suggesting that radial and orbital excitations compete differently in open-flavor systems due to the presence of light quarks and isospin-dependent interactions.

These sizable splittings provide clear experimental targets. The $\Sigma_g^+(1P)$ multiplet ($J^{P}=0^{-},1^{-},2^{-}$) could be searched for in processes such as $B^+ \to D^- D_s^+ \pi^+ \pi^-$ or $B^0 \to \bar{D}^0 D_s^+ \pi^- \pi^0$, where the additional pion(s) carry the necessary orbital angular momentum. Similarly, the $\Sigma_g^+(2S)$ and $\Sigma_g^+(1D)$ states may appear in final states with higher invariant masses, possibly overlapping with open-flavor thresholds such as $D_s^*\rho$ and $D^*K^*$ near $3.4\!-\!3.5\;\text{GeV}$.

The predicted level structure offers a testable pattern for future high-statistics studies, and the systematic trends in excitation energies can help distinguish between compact diquark configurations and other structural interpretations.


\section{Conclusion} \label{conclusion}

In this work, we have analyzed the open-charm tetraquark candidates $T_{c\bar{s}0}^{a}(2900)^{0}$ and $T_{c\bar{s}0}^{a}(2900)^{++}$ within the dynamical diquark model supplemented by the BO approximation. The near-threshold production of these states suggests that the strange quark is not strongly perturbed by short-range dynamics inside the tetraquark, 
permitting its treatment as an effectively heavy, or quasi-heavy, degree of freedom. This motivates the application of BO methods, in which the diquark--antidiquark motion evolves in a confining potential derived from lattice QCD.

Our analysis clearly demonstrates that the internal spin structure of the diquark plays a decisive role in determining the spectroscopy of the $T_{c\bar{s}}$ multiplet.  When the constituent clusters are modeled as scalar ($J^{P}=0^{+}$) diquarks, the predicted masses undershoot the experimental values by approximately 
$150$--$160$~MeV for both charge states. Such a persistent and parameter-stable deficit---substantially larger than any plausible theoretical uncertainty in the diquark masses or in the BO potential---rules out scalar diquarks as the dominant component of the observed resonances.

In contrast, the axial-vector ($J^{P}=1^{+}$) diquark configuration successfully reproduces the experimental spectrum.   The predicted ranges $2881$--$2894$~MeV for $T_{c\bar{s}0}^{a}(2900)^{0}$ and $2940$--$2950$~MeV for $T_{c\bar{s}0}^{a}(2900)^{++}$ closely match the LHCb values once uncertainties are taken into account.  The fact that a single underlying configuration $[cq]_{1^+}\,[\bar{s}\bar{q}]_{1^+}$ describes both charge states without tuning strongly supports the interpretation of these resonances as compact scalar tetraquarks with axial-vector diquark constituents. In this configuration, the axial-vector diquark and antidiquark spins are coupled to an overall $J^P=0^+$ state, consistent with the experimentally determined quantum numbers of the $T^a_{c\bar{s}}$ resonances.

The preference for the axial-vector configuration can also be understood through the effective heavy-quark behavior of the strange quark and the structure of lattice BO potentials. In the near-threshold regime relevant for $T_{c\bar{s}}$ states, the strange quark is not significantly excited by the long-range dynamics of the tetraquark and thus behaves as a quasi-heavy degree of freedom. This suppresses spin-sensitive transitions in the $\bar{s}\bar{q}$ antidiquark and enhances the stability of the axial-vector $[cq]_{1^{+}}$ configuration, whose internal hyperfine structure is more robust against the reduced light-quark spin interactions.  Furthermore, the lattice-determined $\Sigma_g^{+}$ BO potential exhibits a confining structure consistent with a heavy static color source interacting with a lighter partner, a pattern naturally compatible with the intermediate mass scale of the strange quark. The combined effect of quasi-heavy strange-quark dynamics, reduced spin excitations, and the confining BO potential therefore provides a unified explanation for why only the axial-vector diquark configuration reproduces the observed $T_{c\bar{s}}$ spectrum.

The predicted isospin mass splitting, $\Delta M = M^{++} - M^{0} \simeq 56\text{--}62~\mathrm{MeV}$, is stable across all parameter sets, indicating that the splitting arises primarily from the flavor and color--spin structure of the diquark–antidiquark system. The predicted isospin mass splitting,
$\Delta M = M^{++} - M^{0} \simeq 56$--$62~\mathrm{MeV}$, is stable across all parameter sets. Its magnitude is compatible with expectations from
explicit $u$--$d$ mass differences and is consistent with the typical scale of isospin splittings discussed in earlier studies of isovector heavy--light tetraquark systems (see, e.g., Refs.~\cite{Giron:2019cfc,Giron:2020fvd,Giron:2020qpb,Giron:2021sla,Lebed:2023kbm}).

The spatial information extracted from the BO wavefunctions further confirms the compact nature of these states.  
The rms radii of both charge states lie within $0.70$--$0.80$~fm, well below the molecular scale $\gtrsim 1$~fm, decisively disfavoring extended $D^{(*)}K^{(*)}$ configurations. These radii are instead compatible with short-range confinement and the formation of a color flux tube between the diquark and antidiquark.

A definitive confirmation of the isospin assignment would come from the observation of the missing $I_{3}=0$ partner $T^{a+}_{c\bar{s}0}$ in channels such as $B^{+} \to \bar{D}^{0} D_{s}^{+} \pi^{0}$ or $B^{0} \to D^{-} D_{s}^{+} \pi^{+}$.
Moreover, the decay patterns differ for $I=0$ and $I=1$ states: an $I=0$ $c\bar{s}q\bar{q}$ tetraquark would decay predominantly to $D_{s}^{(*)} \eta$ or $D^{(*)} K$ final states with comparable strength, while an $I=1$ state would favor $D_{s}^{(*)} \pi$ and $D^{(*)} K^{*}$ modes.
Future amplitude analyses with higher statistics could test these patterns and provide independent verification of the isospin structure.

Taken together, the mass spectrum, isospin structure, and spatial properties paint a coherent and unified picture:  
the $T_{c\bar{s}}^{a}$ resonances are best interpreted as compact diquark--antidiquark tetraquarks dominated by axial-vector diquark components. The dynamical diquark model, augmented with BO dynamics, therefore provides a predictive and internally consistent framework for describing open-flavor exotics and reveals structural patterns that may extend to bottom--charm, charm--strange, and other mixed heavy--light tetraquark systems.

Further theoretical developments---including spin-dependent splittings, BO-channel mixing, and decay-width calculations---will refine this picture.  Experimentally, improved Dalitz analyses of $B^{0}\to\bar{D}^{0}D_{s}^{+}\pi^{-}$ and 
$B^{+}\to D^{-}D_{s}^{+}\pi^{+}$, as well as searches for additional isospin partners, will be essential for 
fully establishing the internal dynamics and flavor structure of these intriguing states.

\bibliography{Tcs-States-Revised}

\begin{thebibliography}{99}
\bibitem{Gell-Mann:1964ewy}
Gell-Mann, Murray.
\newblock {A Schematic Model of Baryons and Mesons}.
\newblock Phys. Lett. \textbf{8}, 214--215 (1964).
\newblock doi: 10.1016/S0031-9163(64)92001-3.

\bibitem{Zweig:1964CERN}
Zweig, G..
\newblock Developments in the Quark Theory of Hadrons.
\newblock CERN Report No.8182/TH.401, CERN Report No.8419/TH.412  (1964).

\bibitem{Belle:2003nnu}
Choi, S. K. and others.
\newblock {Observation of a narrow charmonium-like state in exclusive $B^\pm \to K^\pm \pi^+ \pi^- J/\psi$ decays}.
\newblock Phys. Rev. Lett. \textbf{91}, 262001 (2003).
\newblock doi: 10.1103/PhysRevLett.91.262001.
\newblock arXiv:hep-ex/0309032.

\bibitem{LHCb:2020pxc}
Aaij, Roel and others.
\newblock {Amplitude analysis of the $B^+\to D^+D^-K^+$ decay}.
\newblock Phys. Rev. D \textbf{102}, 112003 (2020).
\newblock doi: 10.1103/PhysRevD.102.112003.
\newblock arXiv:2009.00026 [hep-ex].

\bibitem{LHCb:2020bls}
Aaij, Roel and others.
\newblock {A model-independent study of resonant structure in $B^+\to D^+D^-K^+$ decays}.
\newblock Phys. Rev. Lett. \textbf{125}, 242001 (2020).
\newblock doi: 10.1103/PhysRevLett.125.242001.
\newblock arXiv:2009.00025 [hep-ex].

\bibitem{LHCb:2022lzp}
Aaij, R. and others.
\newblock {Amplitude analysis of $B^0 \to \bar{D}^0 D_s^+ \pi^-$ and $B^+ \to D^- D_s^+ \pi^+$ decays}.
\newblock Phys. Rev. D \textbf{108}(1), 012017 (2023).
\newblock doi: 10.1103/PhysRevD.108.012017.
\newblock arXiv:2212.02717 [hep-ex].

\bibitem{LHCb:2022sfr}
Aaij, R. and others.
\newblock {First Observation of a Doubly Charged Tetraquark and Its Neutral Partner}.
\newblock Phys. Rev. Lett. \textbf{131}(4), 041902 (2023).
\newblock doi: 10.1103/PhysRevLett.131.041902.
\newblock arXiv:2212.02716 [hep-ex].

\bibitem{Chen:2022svh}
Chen, Rui and Huang, Qi.
\newblock From the isovector molecular explanation of the newly observed
                    $T^{a}_{cs0}(2900)^{++}$ to possible charmed--strange molecular pentaquarks.
\newblock (2022).
\newblock arXiv:2208.10196 [hep-ph].

\bibitem{Agaev:2022eyk}
Agaev, S. S., Azizi, K., and Sundu, H..
\newblock {Modeling the resonance $T^{a}_{cs0}(2900)^{++}$ as a hadronic molecule $D^{*+}K^{*+}$}.
\newblock Phys. Rev. D \textbf{107}(9), 094019 (2023).
\newblock doi: 10.1103/PhysRevD.107.094019.
\newblock arXiv:2212.12001 [hep-ph].

\bibitem{Yue:2022mnf}
Yue, Zi-Li, Xiao, Cheng-Jian, and Chen, Dian-Yong.
\newblock Decays of the fully open-flavor state $T_{cs}(2900)^0$
                    in a $D^\ast K^\ast$ molecule scenario.
\newblock Phys. Rev. D \textbf{107}, 034018 (2023).
\newblock doi: 10.1103/PhysRevD.107.034018.
\newblock arXiv:2212.03018 [hep-ph].

\bibitem{Liu:2022hbk}
Liu, Feng-Xiao, Ni, Ru-Hui, Zhong, Xian-Hui, and Zhao, Qiang.
\newblock {Charmed-strange tetraquarks and their decays in the potential quark model}.
\newblock Phys. Rev. D \textbf{107}(9), 096020 (2023).
\newblock doi: 10.1103/PhysRevD.107.096020.
\newblock arXiv:2211.01711 [hep-ph].

\bibitem{Yang:2023evp}
Yang, Xiao-Song, Xin, Qi, and Wang, Zhi-Gang.
\newblock {Analysis of the $T_{c\bar{s}}(2900)$ and related tetraquark states with the QCD sum rules}.
\newblock Int. J. Mod. Phys. A \textbf{38}(11), 2350056 (2023).
\newblock doi: 10.1142/S0217751X23500562.
\newblock arXiv:2302.01718 [hep-ph].

\bibitem{Lian:2023cgs}
Lian, Ding-Kun, Chen, Wei, Chen, Hua-Xing, Dai, Ling-Yun, and Steele, T. G..
\newblock {Strong decays of $T^a_{c{\bar{s}0}}(2900)^{++/0}$ as a fully open-flavor tetraquark state}.
\newblock Eur. Phys. J. C \textbf{84}(1), 1 (2024).
\newblock doi: 10.1140/epjc/s10052-023-12355-4.
\newblock arXiv:2302.01167 [hep-ph].

\bibitem{Wei:2022wtr}
Wei, Jia, Wang, Yi-Heng, An, Chun-Sheng, and Deng, Cheng-Rong.
\newblock Color flux-tube nature of the states $T_{cs}(2900)$ and
                    $T^{a}_{cs}(2900)$.
\newblock Phys. Rev. D \textbf{106}, 096023 (2022).
\newblock doi: 10.1103/PhysRevD.106.096023.
\newblock arXiv:2210.04841 [hep-ph].

\bibitem{Ortega:2023azl}
Ortega, P. G., Entem, D. R., Fernandez, F., and Segovia, J..
\newblock {Novel $T_{cs}$ and $T_{cs}$ candidates in a constituent-quark-model-based meson-meson coupled-channels calculation}.
\newblock Phys. Rev. D \textbf{108}(9), 094035 (2023).
\newblock doi: 10.1103/PhysRevD.108.094035.
\newblock arXiv:2305.14430 [hep-ph].

\bibitem{Ke:2022ocs}
Ke, Hong-Wei, Shi, Yi-Fan, Liu, Xiao-Hai, and Li, Xue-Qian.
\newblock {Possible molecular states of $\bar{D}^*K^* (D^*K^*)$ and new exotic states $X_0(2900)$, $X_1(2900)$, ($T_{cs0}^a(2900)^0$ and $T_{cs0}^a(2900)^{++}$)}.
\newblock Phys. Rev. D \textbf{106}(11), 114032 (2022).
\newblock doi: 10.1103/PhysRevD.106.114032.
\newblock arXiv:2210.06215 [hep-ph].

\bibitem{Agaev:2022duz}
Agaev, S. S., Azizi, K., and Sundu, H..
\newblock {On the structures of new scalar resonances $T_{cs0}^{a}(2900)^{++}$
and $T_{cs0}^{a}(2900)^{0}$}.
\newblock J. Phys. G \textbf{50}(5), 055002 (2023).
\newblock doi: 10.1088/1361-6471/acc41a.
\newblock arXiv:2207.02648 [hep-ph].

\bibitem{Molina:2022jcd}
Molina, R. and Oset, E..
\newblock {$T_{c\bar{s}}(2900)$ as a threshold effect from the interaction of the $D^*K^*$, $D_s^* \rho$ channels}.
\newblock Phys. Rev. D \textbf{107}(5), 056015 (2023).
\newblock doi: 10.1103/PhysRevD.107.056015.
\newblock arXiv:2211.01302 [hep-ph].

\bibitem{Duan:2023lcj}
Duan, Man-Yu, Du, Meng-Lin, Guo, Zhi-Hui, Wang, En, and Chen, Dian-Yong.
\newblock {Coupled-channel $D^\ast K^\ast -D_s^\ast \rho$ interactions and the origin of $T_{c\bar{s}0}(2900)$}.
\newblock Phys. Rev. D \textbf{108}(7), 074006 (2023).
\newblock doi: 10.1103/PhysRevD.108.074006.
\newblock arXiv:2307.04092 [hep-ph].

\bibitem{Duan:2023qsg}
Duan, Man-Yu, Wang, En, and Chen, Dian-Yong.
\newblock {Searching for the open flavor tetraquark $T_{c\bar{s}0}(2900)^{++}$ in the process $B^+\rightarrow K^+ D^+ D^-$}.
\newblock Eur. Phys. J. C \textbf{84}(7), 681 (2024).
\newblock doi: 10.1140/epjc/s10052-024-13044-6.
\newblock arXiv:2305.09436 [hep-ph].

\bibitem{Wang:2023hpp}
Wang, Bo, Chen, Kan, Meng, Lu, and Zhu, Shi-Lin.
\newblock {Spectrum of the molecular tetraquarks: Unraveling the $T_{cs0}(2900)$ and $T_{c \bar{s}}^a(2900)$}.
\newblock Phys. Rev. D \textbf{109}(3), 034027 (2024).
\newblock doi: 10.1103/PhysRevD.109.034027.
\newblock arXiv:2309.02191 [hep-ph].

\bibitem{Lyu:2023ppb}
Lyu, Wen-Tao, Lyu, Yun-He, Duan, Man-Yu, Li, De-Min, Chen, Dian-Yong, and Wang, En.
\newblock {Roles of the  $T_{c\bar{s}0}(2900)^0$  and $D_0^*(2300)$ in the process $B^- \to D_s^+ K^- \pi^-$}.
\newblock Phys. Rev. D \textbf{109}(1), 014008 (2024).
\newblock doi: 10.1103/PhysRevD.109.014008.
\newblock arXiv:2306.16101 [hep-ph].

\bibitem{Gordillo:2025caj}
Gordillo, M. C. and Segovia, J..
\newblock {Diffusion Monte Carlo calculation of compact $T_{cs0}$ and $T_{c\bar{s}0}$ tetraquarks}.
\newblock Phys. Lett. B \textbf{870}, 139927 (2025).
\newblock doi: 10.1016/j.physletb.2025.139927.
\newblock arXiv:2507.10346 [hep-ph].

\bibitem{Yu:2025xip}
Yu, Zhuo, Wu, Qi, Yue, Zi-Li, and Chen, Dian-Yong.
\newblock {$T_{\bar{c}\bar{s}1}^0$ production in the $B^+$ decays processes}.
\newblock (2025).
\newblock arXiv:2511.18072 [hep-ph].

\bibitem{LHCb:2024vfz}
Aaij, Roel and others.
\newblock {Observation of New Charmonium or Charmoniumlike States in $B^+ \to D^{*\pm} D^{\mp}$ Decays}.
\newblock Phys. Rev. Lett. \textbf{133}(13), 131902 (2024).
\newblock doi: 10.1103/PhysRevLett.133.131902.
\newblock arXiv:2406.03156 [hep-ex].

\bibitem{Born:1927rpw}
Born, M. and Oppenheimer, R..
\newblock {Zur Quantentheorie der Molekeln}.
\newblock Annalen Phys. \textbf{389}(20), 457--484 (1927).
\newblock doi: 10.1002/andp.19273892002.

\bibitem{Braaten:2014qka}
Braaten, Eric, Langmack, Christian, and Smith, D. Hudson.
\newblock {Born-Oppenheimer Approximation for the XYZ Mesons}.
\newblock Phys. Rev. D \textbf{90}(1), 014044 (2014).
\newblock doi: 10.1103/PhysRevD.90.014044.
\newblock arXiv:1402.0438 [hep-ph].

\bibitem{Juge:1999ie}
Juge, K. J., Kuti, J., and Morningstar, C. J..
\newblock {Ab initio study of hybrid anti-b g b mesons}.
\newblock Phys. Rev. Lett. \textbf{82}, 4400--4403 (1999).
\newblock doi: 10.1103/PhysRevLett.82.4400.
\newblock arXiv:hep-ph/9902336.

\bibitem{Brambilla:2017uyf}
Brambilla, Nora, Krein, Gast{\~a}o, Tarr{\'u}s Castell{\`a}, Jaume, and Vairo, Antonio.
\newblock {Born-Oppenheimer approximation in an effective field theory language}.
\newblock Phys. Rev. D \textbf{97}(1), 016016 (2018).
\newblock doi: 10.1103/PhysRevD.97.016016.
\newblock arXiv:1707.09647 [hep-ph].

\bibitem{Berwein:2024ztx}
Berwein, Matthias, Brambilla, Nora, Mohapatra, Abhishek, and Vairo, Antonio.
\newblock {Hybrids, tetraquarks, pentaquarks, doubly heavy baryons, and quarkonia in Born-Oppenheimer effective theory}.
\newblock Phys. Rev. D \textbf{110}(9), 094040 (2024).
\newblock doi: 10.1103/PhysRevD.110.094040.
\newblock arXiv:2408.04719 [hep-ph].

\bibitem{Brambilla:2024imu}
Brambilla, Nora, Mohapatra, Abhishek, Scirpa, Tommaso, and Vairo, Antonio.
\newblock {Nature of $\chi_{c1}(3872)$ and $T_{cc}^+(3875)$}.
\newblock Phys. Rev. Lett. \textbf{135}(13), 131902 (2025).
\newblock doi: 10.1103/pdy7-hvg7.
\newblock arXiv:2411.14306 [hep-ph].

\bibitem{Brambilla:2025xma}
Brambilla, Nora, Mohapatra, Abhishek, and Vairo, Antonio.
\newblock {Unraveling pentaquarks with the Born-Oppenheimer effective theory}.
\newblock Phys. Rev. D \textbf{112}(11), 114037 (2025).
\newblock doi: 10.1103/5z3t-rq5f.
\newblock arXiv:2508.13050 [hep-ph].

\bibitem{Allaman:2024vwn}
Allaman, H{\'e}lo{\"\i}se, Ekhterachian, Majid, Nardi, Filippo, Rattazzi, Riccardo, and Stelzl, Stefan.
\newblock {Tetraquarks at large M and large N}.
\newblock JHEP \textbf{11}, 034 (2024).
\newblock doi: 10.1007/JHEP11(2024)034.
\newblock arXiv:2407.18298 [hep-ph].

\bibitem{Maiani:2022qze}
Maiani, Luciano, Pilloni, Alessandro, Polosa, Antonio D., and Riquer, Veronica.
\newblock {Doubly heavy tetraquarks in the Born-Oppenheimer approximation}.
\newblock Phys. Lett. B \textbf{836}, 137624 (2023).
\newblock doi: 10.1016/j.physletb.2022.137624.
\newblock arXiv:2208.02730 [hep-ph].

\bibitem{LHCb:2021vvq}
Aaij, Roel and others.
\newblock {Observation of an exotic narrow doubly charmed tetraquark}.
\newblock Nature Phys. \textbf{18}(7), 751--754 (2022).
\newblock doi: 10.1038/s41567-022-01614-y.
\newblock arXiv:2109.01038 [hep-ex].

\bibitem{LHCb:2021auc}
Aaij, Roel and others.
\newblock {Study of the doubly charmed tetraquark $T_{cc}^{+}$}.
\newblock Nature Commun. \textbf{13}(1), 3351 (2022).
\newblock doi: 10.1038/s41467-022-30206-w.
\newblock arXiv:2109.01056 [hep-ex].

\bibitem{Ni:2021pce}
Ni, Ru-Hui, Li, Qi, and Zhong, Xian-Hui.
\newblock {Mass spectra and strong decays of charmed and charmed-strange mesons}.
\newblock Phys. Rev. D \textbf{105}(5), 056006 (2022).
\newblock doi: 10.1103/PhysRevD.105.056006.
\newblock arXiv:2110.05024 [hep-ph].

\bibitem{Chen:2023qlx}
Chen, Rui and Huang, Qi.
\newblock {Possible open charm molecular pentaquarks from $\Lambda_c K^{(*)}/ \Sigma_c K^{(*)} $ interactions}.
\newblock Phys. Rev. D \textbf{108}(5), 054011 (2023).
\newblock doi: 10.1103/PhysRevD.108.054011.
\newblock arXiv:2307.04168 [hep-ph].

\bibitem{Silvestre-Brac:1996myf}
Silvestre-Brac, B..
\newblock {Spectrum and static properties of heavy baryons}.
\newblock Few Body Syst. \textbf{20}, 1--25 (1996).
\newblock doi: 10.1007/s006010050028.

\bibitem{Juge:2002br}
Juge, K. Jimmy, Kuti, Julius, and Morningstar, Colin.
\newblock {Fine structure of the QCD string spectrum}.
\newblock Phys. Rev. Lett. \textbf{90}, 161601 (2003).
\newblock doi: 10.1103/PhysRevLett.90.161601.
\newblock arXiv:hep-lat/0207004.

\bibitem{Braaten:2013boa}
Braaten, Eric.
\newblock {How the $Z_c$(3900) Reveals the Spectra of Quarkonium Hybrid and Tetraquark Mesons}.
\newblock Phys. Rev. Lett. \textbf{111}, 162003 (2013).
\newblock doi: 10.1103/PhysRevLett.111.162003.
\newblock arXiv:1305.6905 [hep-ph].

\bibitem{Jafarzade:2025qvx}
Jafarzade, Shahriyar and Lebed, Richard F..
\newblock {Hidden-strangeness tetraquarks in the dynamical diquark model}.
\newblock Phys. Rev. D \textbf{112}(1), 014034 (2025).
\newblock doi: 10.1103/4t78-3tc3.
\newblock arXiv:2505.15704 [hep-ph].

\bibitem{Brodsky:2014xia}
Brodsky, Stanley J., Hwang, Dae Sung, and Lebed, Richard F..
\newblock {Dynamical Picture for the Formation and Decay of the Exotic XYZ Mesons}.
\newblock Phys. Rev. Lett. \textbf{113}(11), 112001 (2014).
\newblock doi: 10.1103/PhysRevLett.113.112001.
\newblock arXiv:1406.7281 [hep-ph].

\bibitem{Giron:2019bcs}
Giron, Jesse F., Lebed, Richard F., and Peterson, Curtis T..
\newblock {The Dynamical Diquark Model: First Numerical Results}.
\newblock JHEP \textbf{05}, 061 (2019).
\newblock doi: 10.1007/JHEP05(2019)061.
\newblock arXiv:1903.04551 [hep-ph].

\bibitem{Giron:2019cfc}
Giron, Jesse F., Lebed, Richard F., and Peterson, Curtis T..
\newblock {The Dynamical Diquark Model: Fine Structure and Isospin}.
\newblock JHEP \textbf{01}, 124 (2020).
\newblock doi: 10.1007/JHEP01(2020)124.
\newblock arXiv:1907.08546 [hep-ph].

\bibitem{Giron:2020fvd}
Giron, Jesse F. and Lebed, Richard F..
\newblock {Spectrum of $p$-wave hidden-charm exotic mesons in the diquark model}.
\newblock Phys. Rev. D \textbf{101}(7), 074032 (2020).
\newblock doi: 10.1103/PhysRevD.101.074032.
\newblock arXiv:2003.02802 [hep-ph].

\bibitem{Giron:2020qpb}
Giron, Jesse F. and Lebed, Richard F..
\newblock {Spectrum of the hidden-bottom and the hidden-charm-strange exotics in the dynamical diquark model}.
\newblock Phys. Rev. D \textbf{102}(1), 014036 (2020).
\newblock doi: 10.1103/PhysRevD.102.014036.
\newblock arXiv:2005.07100 [hep-ph].

\bibitem{Giron:2021sla}
Giron, Jesse F., Lebed, Richard F., and Martinez, Steven R..
\newblock {Spectrum of hidden-charm, open-strange exotics in the dynamical diquark model}.
\newblock Phys. Rev. D \textbf{104}(5), 054001 (2021).
\newblock doi: 10.1103/PhysRevD.104.054001.
\newblock arXiv:2106.05883 [hep-ph].

\bibitem{Mutuk:2022nkw}
Mutuk, Halil.
\newblock {Spectrum of $cc \bar{b}\bar{b}$, $bc\bar{c}\bar{c}$, and $bc\bar{b}\bar{b}$ tetraquark states in the dynamical diquark model}.
\newblock Phys. Lett. B \textbf{834}, 137404 (2022).
\newblock doi: 10.1016/j.physletb.2022.137404.
\newblock arXiv:2208.11048 [hep-ph].

\bibitem{Lebed:2023kbm}
Lebed, Richard F. and Martinez, Steven R..
\newblock {Exotic hadrons from scattering in the diabatic dynamical diquark model}.
\newblock Phys. Rev. D \textbf{108}(1), 014013 (2023).
\newblock doi: 10.1103/PhysRevD.108.014013.
\newblock arXiv:2305.09146 [hep-ph].

\bibitem{Lebed:2017min}
Lebed, Richard F..
\newblock {Spectroscopy of Exotic Hadrons Formed from Dynamical Diquarks}.
\newblock Phys. Rev. D \textbf{96}(11), 116003 (2017).
\newblock doi: 10.1103/PhysRevD.96.116003.
\newblock arXiv:1709.06097 [hep-ph].

\bibitem{Wang:2010sh}
Wang, Zhi-Gang.
\newblock {Analysis of the scalar and axial-vector heavy diquark states with QCD sum rules}.
\newblock Eur. Phys. J. C \textbf{71}, 1524 (2011).
\newblock doi: 10.1140/epjc/s10052-010-1524-y.
\newblock arXiv:1008.4449 [hep-ph].

\bibitem{Wang:2011ab}
Wang, Zhi-Gang.
\newblock {Analysis of the light-flavor scalar and axial-vector diquark states with QCD sum rules}.
\newblock Commun. Theor. Phys. \textbf{59}, 451--456 (2013).
\newblock doi: 10.1088/0253-6102/59/4/11.
\newblock arXiv:1112.5910 [hep-ph].

\bibitem{Maiani:2004vq}
Maiani, L., Piccinini, F., Polosa, A. D., and Riquer, V..
\newblock {Diquark-antidiquarks with hidden or open charm and the nature of X(3872)}.
\newblock Phys. Rev. D \textbf{71}, 014028 (2005).
\newblock doi: 10.1103/PhysRevD.71.014028.
\newblock arXiv:hep-ph/0412098.

\bibitem{Giron:2020wpx}
Giron, Jesse F. and Lebed, Richard F..
\newblock {Simple spectrum of $c\bar c c\bar c$ states in the dynamical diquark model}.
\newblock Phys. Rev. D \textbf{102}(7), 074003 (2020).
\newblock doi: 10.1103/PhysRevD.102.074003.
\newblock arXiv:2008.01631 [hep-ph].

\end{thebibliography}

\end{document}